\documentstyle[12pt]{article}

\def\babar{{\sc BaBar}}
\def\dirc{{\sc Dirc}}
\newcommand{\cer} {\v{C}erenkov}

\begin{document}

%%%%%%%%%%%%%%%
%% The following lines create the SLAC Pub Title Page
%%%%%%%%%%%%%%%

\pagestyle{empty}

\renewcommand{\thefootnote}{\fnsymbol{footnote}}

%%%%%%%%%%%%%%%
%% Substitute your Pub number, month and year
%% in the following
%%%%%%%%%%%%%%%

\begin{flushright}
{\small
SLAC--PUB--7602\\
July 1997\\}
\end{flushright}

\vspace{1.cm}

\begin{center}
{\large\bf
D{\normalsize\bf IRC}, the Internally Reflecting Ring Imaging \\
{\v{C}erenkov} Detector 
for B{\normalsize\bf A}B{\normalsize\bf AR}: \\[1mm]
Properties of the Quartz Radiators\footnote{Work supported by
Department of Energy contract  DE--AC03--76SF00515.}}

\vspace{1.cm}

{\bf Jochen Schwiening}

Stanford Linear Accelerator Center

Stanford University, Stanford, CA 94309

\vspace{3.mm}
{\it Representing}
\vspace{.2cm}

{\bf The B{\small\bf A}B{\small\bf AR} D{\small\bf IRC} Collaboration}

\end{center}

\vfill

%%%%%%%%%%%%%%%
%% Choose"Presented at," "Contributed to" for conference papers
%% or "Submitted to" for journal papers
%%%%%%%%%%%%%%%
\begin{center} 
{\it Contributed to the}\\
{\it 7th ICFA School on Instrumentation in Elementary Particle Physics}\\
{\it Le{\'o}n, Guanajuato, M{\'e}xico}\\
{\it July 7-19, 1997 }\\
\end{center}

\vfill\eject

%%%%%%%%%%%%%%%
%% Following are the commands to create the rest
%%of the SLAC Pub.
%%%%%%%%%%%%%%%

%%%%%%%%%%%%%%%
%% The next two lines change the line spacing to doublespace
%%%%%%%%%%%%%%%

%\renewcommand{\baselinestretch}{2}
\renewcommand{\baselinestretch}{1.2}
\normalsize

%%%%%%%%%%%%%%%
%% Paper starts here
%%%%%%%%%%%%%%%

\begin{center}
{\bf D{\small\bf IRC}, the Internally Reflecting Ring \\
Imaging {\v{C}erenkov} Detector 
for B{\small\bf A}B{\small\bf AR}: \\
Properties of the Quartz Radiators}
\end{center}

A new type of detector for particle identification will be used in the {\babar}
experiment~\cite{tdr} at the SLAC B Factory ({\sc Pep-II})~\cite{pep2}. 
This barrel region detector is called {\dirc}, an acronym for 
{\bf D}etection of {\bf I}nternally {\bf R}eflected {\bf{\v{C}}}erenkov 
(light). 
The \dirc\ is a \cer\ ring imaging device
which utilizes totally internally reflecting \cer\ photons
in the visible and near UV range \cite{dirc_concept}.
It is thin (in both size and radiation length), robust and very fast.
An extensive prototype program, progressing  
through a number of different prototypes
in a hardened cosmic muons setup at SLAC~\cite{hide_ieee}
and later on in a test beam at CERN~\cite{nim_paper}, 
demonstrated that the principles of operation are well understood,
and that an excellent performance over the entire momentum range of the 
B factory is to be expected.

The DIRC utilizes long, thin, flat quartz radiator bars
(effective mean refractive index $n_{1}$ = 1.474)
with a rectangular cross section.
The quartz bar is surrounded by a material
with a small refractive index $n_{3} \sim 1$ (nitrogen in this case).
For particles with \mbox{$\beta$ = 1},
%since the index of the radiator bar $n_{1}$ is larger than $\sqrt{2}$,
some of the \cer\ photons will be totally internally reflected,
regardless of the incidence angle of the tracks, and propagate 
along the length of the bar. 
To avoid having to instrument both bar
ends with photon detectors, a mirror is placed at one
end, perpendicular to the bar axis.  
This mirror returns most
of the incident photons to the other (instrumented) bar end.
Since the bar has a rectangular cross section,
the direction of the photons remains unchanged during the transport,
except for left-right/up-down ambiguities due to the reflection
at the radiator bar surfaces.
The photons are then proximity focused by expanding through
a stand-off region filled with purified water (index $n_{2} \sim 1.34$)
onto an array of densely packed photomultiplier tubes 
placed at a distance of about 1.2 m from the bar end.
In the present design the bars have transverse dimensions of 1.7 cm thick 
by 3.5 cm wide, and are about 4.90 m long. 
The length is achieved by gluing end-to-end four
1.225 m bars, that size being the longest high
quality quartz bar  currently available from industry.

Several natural and synthetic fused silica candidate materials 
were tested for their optical properties and radiation hardness.
In a Co$^{60}$ source, samples were exposed to doses of up to 500 krad.
While natural quartz materials showed significant absorbtion in the 
wavelength range of the {\cer} photons 
after being exposed to only a few krad, the synthetic material 
proved to be sufficiently radiation hard.
This led to the choice of 
Suprasil~Standard~\cite{heraeus} and Spectrosil~2000~\cite{qpc} 
as bar material for the \dirc .

Bars were formed from the synthetic quartz material, produced as large
cylindrical ingots, using modifications of conventional optical processing
techniques \cite{zygo}.  
In order to preserve the photon angles during surface reflections,
the faces and sides were
nominally parallel while the orthogonal surfaces were kept nominally
perpendicular. 
Typically, the bar's surfaces were flat and parallel to
about 25 $\mu$m, while the orthogonal surfaces were perpendicular to a
tolerance of 0.3 mrad. 
The most difficult requirements were associated with maintaining the photon 
transmission during reflections at the surfaces of the bar (a \cer\ photon
may be internally reflected a few hundred times before exiting the
bar).  
This led to rather severe requirements on edge sharpness and surface finish.
After polishing, the bars had an average edge radius less than
5 $\mu$m, and a nominal surface polish of better than 0.5 nm (RMS).
The optical properties of the radiator bars were measured 
using a HeCd laser in a motion-controlled setup.
The absorption of a quartz bar is typically about 1\%/m at 325 nm and 
less than 0.2\%/m at 442 nm.
The coefficient of total internal reflection at 442 nm was found to be 
(0.99960 $\pm$ 0.00006), consistent with the expected reflectivity 
for the nominal surface polish.

%%%%%

\end{document}